\documentclass[pre,twoside,showpacs,twocolumn,floatfix]{revtex4}

\usepackage{amssymb}
\usepackage{amsfonts}
\usepackage{amsmath}
\usepackage[dvips]{graphicx}
\usepackage{subfigure}

\usepackage{soul}
\usepackage{color}

\graphicspath{{figures/}}
\begin{document}

\title{Hole dynamics in vertically vibrated liquids and suspensions}
\author{Stefan von Kann, Matthias van de Raa, and Devaraj van der Meer}
\affiliation{Physics of Fluids group, University of Twente, P.O.
Box 217, 7500 AE Enschede, The Netherlands}

\date{\today}

\begin{abstract}
We study the dynamics of holes created in vertically vibrated dense suspensions and viscous Newtonian liquids. We find that all holes oscillate with the driving frequency, with a phase shift of $\pi/2$. In Newtonian liquids holes always close, while in suspensions holes may grow in time. We present a lubrication model for the closure of holes which is in good agreement with the experiments in Newtonian liquids. The growth rate of growing holes in suspensions is found to scale with the particle diameter over the suspending liquid viscosity.
Comparing closing holes in Newtonian liquids to growing holes in dense suspensions we find a sinusoidal, linear response in the first, and a highly non-linear one in the latter. Moreover, the symmetry of the oscillation is broken and is shown to provide an explanation for the observation that holes in dense suspensions can grow.
\end{abstract}

\pacs{82.70.Kj, 45.70.Vn, 47.50.-d}

\maketitle

\section{Introduction} \label{introduction}

When a hole is created in a horizontal layer of (viscous) liquid at rest, the hydrostatic pressure will cause the hole to close. And, in spite of its more complicated rheology, the same thing is expected to happen in a non-Newtonian liquid. Recently however, the reverse has been shown to occur in experiments where layers of various particulate suspensions and emulsions were subjected to vertical vibrations: Holes created in these vibrated liquids do not necessarily close, but may stabilize~\cite{refMerkt04,reffalcon12}, grow~\cite{refEbata09}, or lead to chaotic dynamics~\cite{refEbata11,refKann12}. Although phenomenological models are suggested in the literature~\cite{refEbata09,refDeegan10} our understanding of this behavior is far from complete. In this paper we will shed light onto its dynamics by investigating the analogies and differences between vertically vibrated viscous Newtonian fluids and a suspension of monodisperse particles in liquids with the viscosity of water and higher.

A concentrated particulate suspension consists of a mixture of a homogeneous liquid and particles that are large enough ($> 1\mu m$) such that their Brownian motion is negligible. They can be found in many places, ranging from quicksand, through freshly mixed cement and paints to the inside of flexible armor suits. Their flow is important in nature, industry and even health care~\cite{refWagner09}. In spite of their common presence and significance, many aspects of the flow of these dense suspensions remain poorly understood. In order to study these materials people have used methods inspired by classical rheology, and typically characterized them in terms of a constitutive relation of stress versus shear rate~\cite{refBarnes89,refFall08,refBrownJaeger09,refBrown10,refBonnoit10,refBrown12}. A general result is that, when increasing the shear rate, dense suspensions first tend to become less viscous (shear thinning) and subsequently shear thicken. In recent experiments people found mesoscopic length scales~\cite{refBonnoit10,refBonnoit10prl}, fracturing~\cite{refWhite10}, and a dynamic jamming point~\cite{refBrownJaeger09} to be important in such suspensions. Connected to the above, normal stress divergence in the approach to a wall~\cite{refLiu10} and non-monotonic settling~\cite{refKann11} have been reported for objects moving through dense cornstarch suspensions.

Turning to vertically vibrated suspensions, Merkt~\textit{et al.}~\cite{refMerkt04} observed in a vertically shaken, thin layer of cornstarch suspension that --amongst other quite exotic phenomena-- stable oscillating holes can be formed for certain values of the shaking parameters~\cite{refMerkt04}. These stable holes were subsequently described using a phenomenological model based on a hysteretic constitutive equation~\cite{refDeegan10}. In other particulate suspensions, Ebata~\textit{et al.} found growing and splitting holes and a separated state~\cite{refEbata09,refEbata11}, where the latter is attributed to a convective flow in the rim and the first are still not understood. Stable holes and kinks (which appear to be similar to or even identical to the separated state mentioned above) have also been reported in emulsions~\cite{reffalcon12}. At present we are still far from a detailed understanding of dense suspensions, and why different suspension behave differently.

Here, we will investigate the dynamics of opening holes in a layer of vibrated suspension of monodisperse particles of various sizes suspended in a glycerol-water mixture. We will investigate how this dynamics depends on particle size and viscosity and will compare it to the dynamics of closing holes in a layer of vertically shaken viscous Newtonian liquids, for which we will present a model within the lubrication approximation. We will then shed light upon how the differences arise and in what manner these can explain the observation that the holes in the suspension do not close as a result of hydrostatic pressure.

The paper is organized as follows: We will start with a short description of our setup in Section~\ref{setup}. After this we will present experiments for the dynamics of closing holes in a (vibrated) layer of a viscous Newtonian liquid (Section~\ref{liquids}), followed by the introduction and discussion of a lubrication model for this system (Section~\ref{model}). Subsequently, in Section~\ref{suspensions} we turn to the dynamics of opening holes in vibrated particle suspensions and discuss the similarities and differences with the closing holes. The paper will be concluded in Section~\ref{conclusions}.

\section{Experimental setup} \label{setup}

\begin{figure}[tbp]
\centering
\includegraphics[width=0.8\linewidth]{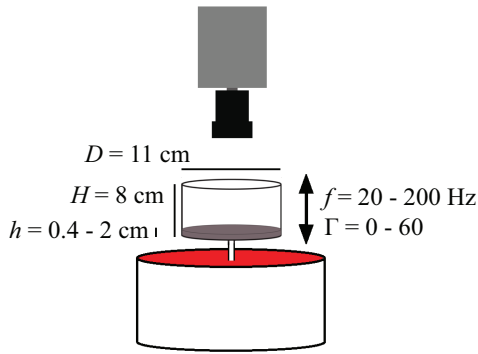}
\caption{A schematic view of the used setup. At the lower end we have the shaker, on top of which with the container with the suspension is mounted, which is subsequently vibrated vertically. Above that is the high speed camera, recording the suspension from above.}
\label{pic:setup}
\end{figure}

The experimental setup is shown in Fig.~\ref{pic:setup}. The core consists of a cylindrical container with a diameter $D = 11.0$ cm and a height $H$ of $8.0$ cm. This container is vertically vibrated by a shaker (TiraVib 50301) with frequencies $f$ between $20$ and $200$ Hz and a dimensionless acceleration $\Gamma$ from 0 up to 60. Here, $\Gamma= a (2 \pi f)^2/g$, where $a$ is the shaking amplitude and $g$ the gravitational acceleration. The container is filled up to a height $h = 6.0 \pm 0.5$ mm with a viscous liquid or a suspension of varying composition.
The dynamics of the fluid layer in the container is recorded with a high speed camera at frame rates ranging from $500$ to $5,000$ frames per second (fps), and is imaged from the top. The bottom of the container was covered with an adhesive sheet for improved contrast between liquid and container bottom. When using transparent liquids, a small amount of powdered milk was added to whiten the liquid. Of course it was checked that adding adhesive sheet or milk powder did not influence the dynamics of the system.

\section{Viscous Newtonian liquids}
\label{Newtonian}

Before turning to the --anomalous-- opening holes in dense suspensions consisting of monodisperse particles in a mixture of glycerine and water, we will first study the regular case of holes closing in a viscous Newtonian liquid. We will both discuss the case where the holes close purely due to the hydrostatic pressure in the liquid and the case in which a periodic forcing is added by vibrating the system vertically. In the second subsection we will subsequently present a model to describe both cases.

\subsection{Experiment}\label{liquids}
We prepare a layer with a thickness of $h=6.0\pm0.5$ mm of a viscous liquid in the container as described in the previous Section. Subsequently, a disturbance is created into the layer by blowing air from the top until an approximately circular hole with a diameter of a few centimeters is formed.
To vary the viscosity of the liquid we choose honey, with a dynamic viscosity of $\mu = 6.4$ Pa$\cdotp$s, and several glycerine-water mixtures with viscosities of $\mu = 1.30$, $1.10$, $0.45$, and $0.15$ Pa$\cdotp$s. Viscosities below the last value lead to holes that close extremely fast; in particular they were found to close within a single cycle of the lowest driving frequency we have used in our study ($f = 20$ Hz). Moreover, for these low viscosities inertial effects will start to become important and therefore such fluids were not considered here.

\begin{figure}[tbp]
\centering
\includegraphics[width=0.99\linewidth]{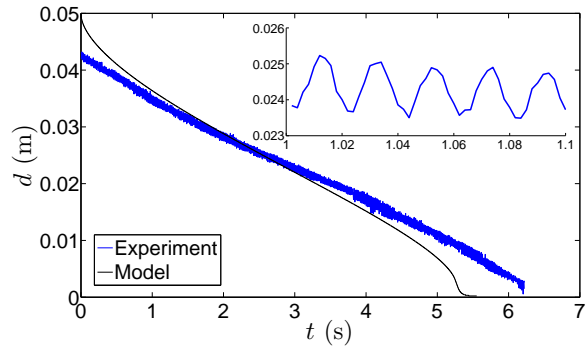}
\caption{The diameter of a closing hole in a layer of honey ($\mu=6.4$ Pa$\cdotp$s) with a thickness of $h = 6 \pm 1$ mm  as a function of time, while the layer is vertically vibrated at $f=50$ Hz, $\Gamma=30$ and recorded at a framerate of $250$ fps (blue line). The black line is the result of a calculation using the lubrication model. The inset is from an experiment using the same shaking parameters, but twice the recording speed ($500$ fps).}
\label{pic:honey}
\end{figure}

Fig.~\ref{pic:honey} provides a typical experimental result for a $h=6$ mm thick layer of honey, vibrated at $f=50$ Hz, $\Gamma=30$. After creating a circular hole in the layer, we follow the dynamics of its closing and plot the hole diameter as a function of time. Over the course of several seconds --i.e., on time scales larger than the period of the driving-- the hole is observed to close in an almost linear manner. On top of this the hole oscillates at the same frequency as that of the driving, which is shown in the inset where part of the signal has been magnified in time.

\begin{figure}[tbp]
\centering
\includegraphics[width=0.99\linewidth]{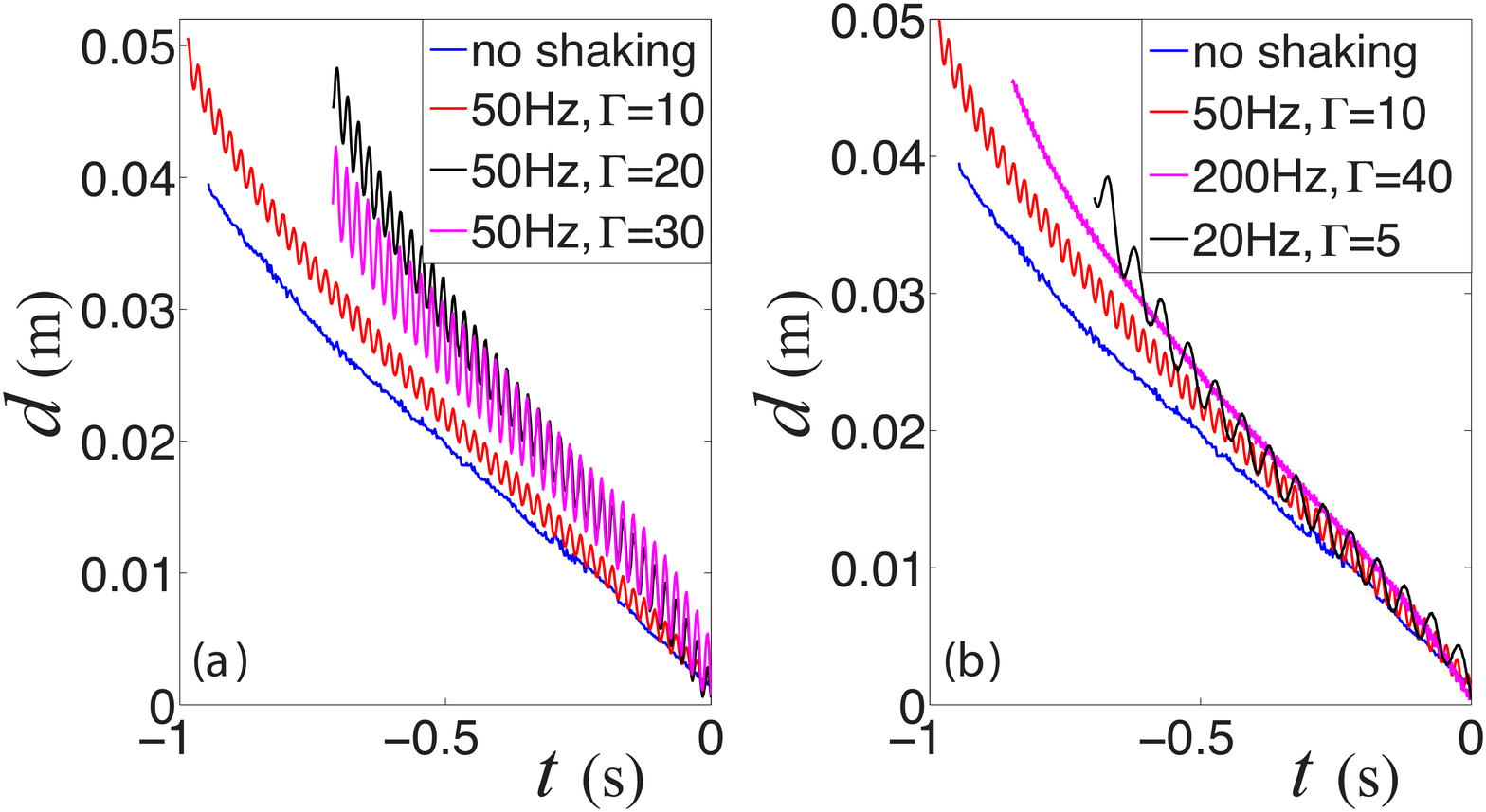}
\caption{Time evolution of the hole diameter in glycerine ($\mu=1.3$ Pa$\cdotp$s): (a) For a constant shaking frequency $f=50$ Hz and different values of the shaking acceleration $\Gamma=0$ (no shaking, blue line), $\Gamma=10$ (red line), $\Gamma = 20$ (magenta line), and $\Gamma = 30$ (black line). (b) For varying frequency: $f=0$ Hz, $\Gamma=0$ [no shaking, blue line, as in (a)]; $f=20$ Hz, $\Gamma=5$ (black line); $f=50$ Hz, $\Gamma=10$ [red line, as in (a)]; and $f=200$ Hz, $\Gamma=40$ (magenta line).}
\label{pic:Nvarious}
\end{figure}

When 
changing the shaking parameters $f$ and $\Gamma$, it becomes clear that the closing time is to a large extent independent on $f$ and $\Gamma$, as is shown in Fig.~\ref{pic:Nvarious} where we show results obtained in glycerine. In particular, when we do not shake at all and just create a hole in the container at rest and observe its closing, we find that its time evolution follows the very same trend. The amplitude of the oscillation increases more or less linearly with the shaking acceleration $\Gamma$ and is in fact of the same order as the shaking amplitude $a = \Gamma g/(2\pi f)^2$. The latter observation also explains why the amplitude of the oscillation decreases so much when the frequency is raised to $f=200$ Hz, which causes the shaking amplitude to go down by a factor $16$.  Moreover, the amplitude appears to be independent of the hole size, i.e., the amplitude remains largely constant while the hole diameter shrinks down to zero.

\begin{figure}[tbp]
\centering
\includegraphics[width=0.99\linewidth]{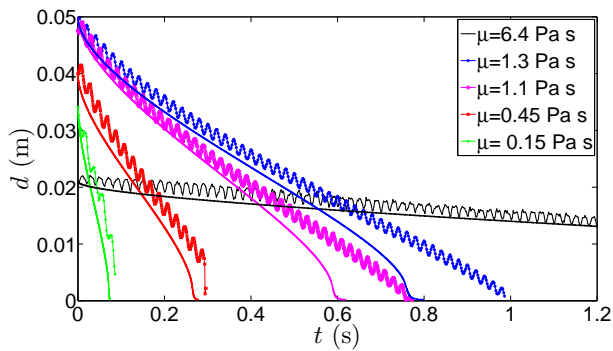}
\caption{Time evolution of the hole diameter in a $h=6$ mm thick layer of liquid of varying viscosity $\mu$, vibrated at $f=50$ Hz and $\Gamma = 10$. The solid black lines denote the time evolution according to the model discussed in Section~\ref{model}.}
\label{pic:expmodelN}
\end{figure}

In Fig.~\ref{pic:expmodelN} we compare results for the different liquid viscosities, shaken at $f=50$ Hz and $\Gamma=10$. For the lowest viscosity ($\mu=0.15$ Pa$\cdotp$s ) we observe that the holes closes in less than a tenth of a second, i.e., within a few cycles of the driving. When we increase the viscosity the closing time increases rapidly, and for the highest viscosity (that of honey, $\mu=6.4$ Pa$\cdotp$s) the closing time is over six seconds.

\begin{figure}[tbp]
\centering
\includegraphics[width=0.99\linewidth]{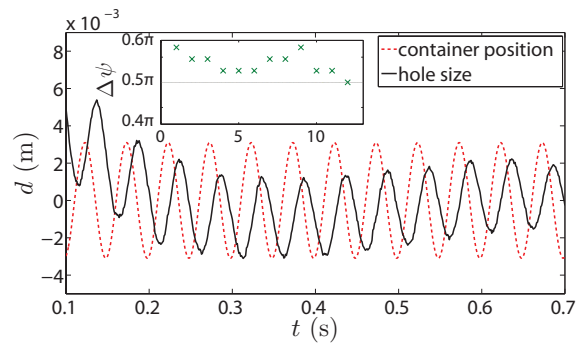}
\caption{Comparison of a sine fit of the trajectory of the vertical position of the container (vibrated at $f=20$ Hz and $\Gamma = 5$), and the trajectory of diameter of a closing hole in a $h=6$ mm layer of glycerine ($\mu = 1.3$ Pa$\cdotp$s), corrected linearly for the average closing velocity of the hole (see text). The inset shows the phase difference $\Delta\phi$ for every period shown in the main Figure.
}
\label{pic:Nphase}
\end{figure}

In the same Figure we observe that there is a significant span of time in which the average closure velocity appears to be linear. This allows us to correct the signal by subtracting this linear behavior and afterwards compare it to the vertical position of the container. This is done in Fig.~\ref{pic:Nphase}, where we zoom in on a few cycles only. 
There is a clear phase shift between the driving and the hole, which is measured to be approximately a quarter of a period, as shown in the inset of Fig.~\ref{pic:Nphase}. The fact that the horizontal oscillation of the hole lags behind $\Delta\phi=\pi/2$ with the vertical container position implies that the oscillating velocity of the hole is in phase with the latter. This in turn implies that the velocity with which the hole oscillates is in antiphase with the shaking acceleration.

\begin{figure}[tbp]
\centering
\includegraphics[width=0.99\linewidth]{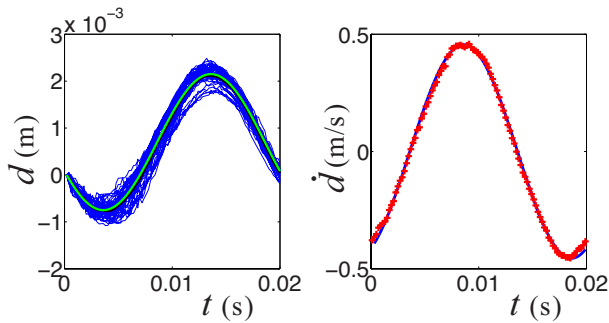}
\caption{(a) Superposition of several cycles of the corrected trajectory of the diameter $d(t)$ of a closing hole shifted over an integer number of periods of the driving. The (almost completely overlapping) black and yellow curves indicate the average position and a sinusoidal fit to the average respectively. (b) The instantaneous velocity $\dot{d}(t)$ of the closing hole averaged over all cycles (red $+$ symbols). The solid (blue) line shows the derivative of the sine fit of (a). Taken from an experiment with glycerine ($\mu = 1.3$ Pa$\cdotp$s, $h=6$mm, $f=50$ Hz, $\Gamma = 20$).
}
\label{pic:Nsymmetry}
\end{figure}

To quantitatively compute the average velocity profile in a cycle we start from the corrected signal and shift all cycles on top of each other, as seen in Fig.~\ref{pic:Nsymmetry}(a). We then compute the average diameter and the average velocity [Fig.~\ref{pic:Nsymmetry}(b)]. We conclude that to within measurement precision both the average signal and velocity are nicely sinusoidal. Most importantly the positive and negative half of each cycle are close to each others mirror images and follow the driving very well.

Now what is happening physically? First one should realize that the magnitude of the shaking acceleration that we subject our liquid layers to is many times that of gravity. This means that the liquid layer is alternately subjected to a large downwards acceleration, that is forcing the hole to close --as gravity does-- in one half of the driving period, and an almost equally large upwards acceleration in the other half. Clearly, in this stage the liquid strives to move upward, and therewith opens the hole again. It is the small unbalance between the upward and the downward acceleration caused by gravity that makes the hole close in the long run \footnote{Of course, when the direction of the acceleration is upward, the fluid surface becomes potentially unstable, which one could call a Rayleigh-Taylor, Richtmyer-Meshkov, or Faraday instability, depending on the perspective and the specific timescale one is looking at. This instability is counteracted by surface tension (which stabilizes the smaller wavelengths) and liquid viscosity. From the experiment we infer that for the liquids in this study this leads to stable standing wave patterns (the Faraday waves) in the worst case.}.

\subsection{Modeling}\label{model}

\begin{figure}[tbp]
\centering
\includegraphics[width=0.99\linewidth]{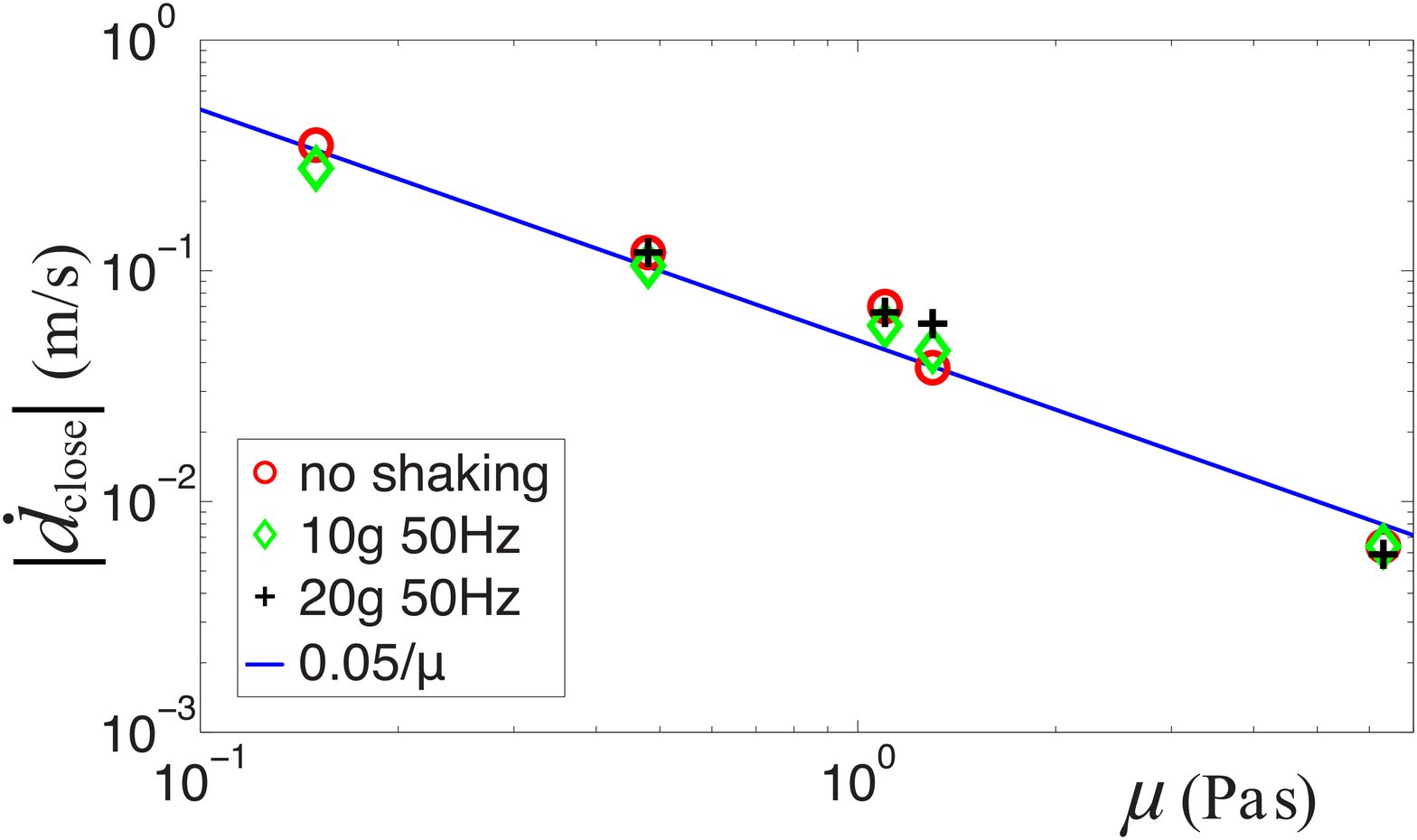}
\caption{Average closing velocity as a function of viscosity in a double logarithmic plot, for three different values for the driving (no shaking; $f=50$ Hz and $\Gamma=10$; and $f=50$ Hz and $\Gamma=20$. The blue solid line is $|\dot{d}_{close}| \approx  0.05/\mu$, making the proportionality constant in Eq.~(\ref{eq:vmu}) equal to $0.14$.}
\label{pic:Nclosingvmu}
\end{figure}

To model the dynamics of closing holes in a viscous Newtonian liquid we use axisymmetric lubrication theory. In absence of the driving, the equation of motion for the liquid profile $h(r,t)$ can be derived from continuity and a lubrication ansatz for the velocity profile within the layer (See Appendix~\ref{appendix} \footnote{Some technical details in the derivation of several results presented in this Section can be found in Appendix~\ref{appendix}.}).
\begin{equation}\label{eq:gravityonly}
\frac{ \partial h}{ \partial t} = \frac{\rho g}{3 \mu r} \frac{\partial}{\partial r} \left[r h^3 \frac{ \partial h}{\partial r} \right]\,,
\end{equation}
where $r$ is the radial coordinate, $g$ the acceleration of gravity, $\rho$ the density, and $\mu$ the dynamic viscosity of the liquid. Using lubrication theory implies neglecting inertial effects. In particular this means that the closing velocity can be derived from an (instantaneous) balance of the gravitational force drives the closing and the viscous forces that counteract it, i.e.
\begin{equation}\label{eq:vmu}
\rho g \sim \mu \frac{\dot{d}}{h_0^2} \quad \Rightarrow \quad \dot{d} \sim  \frac{\rho g h_{0}^{2}}{\mu}\,,
\end{equation}
in which $\dot{d}$ denotes the time derivative of the hole diameter and we have estimated the viscous forces in the layer, $\mu\, \partial^2 u/\partial z^2$ as the velocity of the rim $\dot{d}$ divided by the squared initial layer thickness $h_0$. From this simple balance it follows that the closing velocity should scale as $1/\mu$. If we check this for our experimental results by plotting the closing velocity $\dot{d}$ (determined from the linear regime of plots as in Fig.~\ref{pic:expmodelN}) as a function of viscosity $\mu$ Fig.~\ref{pic:Nclosingvmu} we find a very good agreement. Remarkable is that the plot does not only contain data without driving, but also with various driving strengths.

When we assume an infinite layer of liquid, we can derive a semi-analytical self-similar solution to the closing hole problem which has the form
\begin{equation}\label{eq:selfsim} 
d(t) = 2 \eta_0 \sqrt{\frac{\rho g h_0^3}{3\mu}(t_c-t)\,\,}\,\,,
\end{equation}
where $\eta_0$ is a numerical constant and $t_c$ is the time the hole needs to close. In our case these can be thought of as fixed by the initial hole size together with the boundary conditions at the sidewalls of our container. This self-similar solution goes to zero with a square-root dependence on time which is however --maybe with the exception of the very end-- not observable in our experiments (Figs.~\ref{pic:honey},~\ref{pic:Nvarious} and~\ref{pic:expmodelN}), which is presumably connected to the proximity of the side walls. We therefore decided to numerically solve Eq.~(\ref{eq:gravityonly}), supplemented with $\int_0^{D/2} h(r,t) rdr = \text{constant}$, which expresses the conservation of liquid in our system.

In Figs.~\ref{pic:honey} and~\ref{pic:expmodelN} we compare our model results to the experiments and find that the behavior is well captured by the model.

We can adapt Eq.~(\ref{eq:gravityonly}) to model the modulation due to the acceleration of the shaker as well, by simply substituting $g(1+\Gamma \sin\omega t)$ for $g$, leading to
\begin{equation}\label{eq:full}
\frac{ \partial h}{ \partial t} = (1 + \Gamma \sin\omega t)\,\frac{\rho g}{3 \mu r} \frac{\partial}{\partial r} \left[r h^3 \frac{ \partial h}{\partial r} \right]
\end{equation}
The result is (at least in first order) the same as for the purely gravitational case, with a continuous oscillation on top of the gravitational result, just like we see in our experiments. More details of this calculation can be found in Appendix~\ref{appendix}.

\section{Non-Newtonian liquids}\label{suspensions}

Whereas disturbances created in a layer of a Newtonian liquid always close, independent of whether the layer is being vertically vibrated or not, for non-Newtonian liquids things are observed to be different: More specifically, for the particulate suspensions studied here \footnote{As a yield stress has been reported in some very dense suspensions, it is conceivable that gravity is not capable of overcoming this yield strength when the material is at rest. In the suspensions studied here, this is not the case.}, holes close when the suspensions are at rest, but may either open or close when vertically vibrated.

\subsection{Experiment}\label{suspensionexp}

As discussed in Section~\ref{introduction}, several types of non-closing holes were found in various vibrated suspensions and emulsions, including stable holes, splitting holes, and growing holes~\cite{refMerkt04,refDeegan10,refEbata09,refEbata11,reffalcon12,refKann12}. It is this last type, the growing holes, which will be the focus of this Section. Growing holes are typically found in suspensions containing monodisperse particles \footnote{Stable holes have been observed in cornstarch suspensions and some emulsions~\cite{refMerkt04,reffalcon12}, and splitting holes occur in suspensions of particles with a substantial poydispersity~\cite{refKann12}.}. We therefore use monodisperse, spherical polystyrene particles with a diameter ($\sigma$) of $20$, $40$, and $80$ $\pm$5 $\mu$m, and a density of 1050 kg/m$^3$ (MicroBeads, TS 20-40-80). As the suspending liquid we used various glycerine-water mixtures, with varying viscosities and densities. Because the suspending liquid may be either denser or less dense than the particles, we do not attempt to density match the liquid. In all cases the time scale at which the suspension separates is much larger than the time scales of the experiment. In some cases we have checked that our results did not depend on whether the liquid density would be larger or smaller than that of the particles by adding cesium chloride to the suspending liquid. Much care has been taken to ensure that the packing fraction $\phi$ --the volume of the solid phase in the suspension divided by the total volume-- was kept at a constant value of $0.52$.

\begin{figure}[tbp]
\centering
\includegraphics[width=0.99\linewidth]{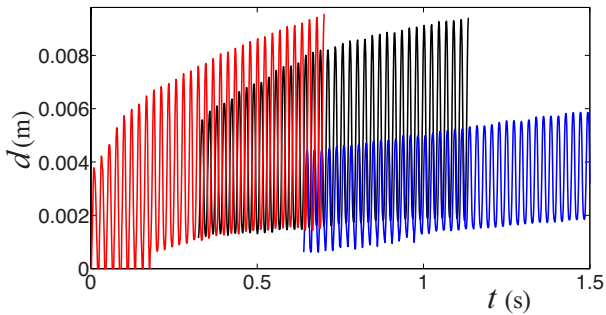}
\caption{Time evolution of the diameter $d$ of a growing hole in a suspension of  $\sigma=40$ $\mu$m polystyrene particles in glycerine-water mixtures of three different viscosities, namely $\mu = 0.14$ Pa$\cdotp$s (red line), $\mu = 0.22$ Pa$\cdotp$s (black line), and $\mu = 0.52$ Pa$\cdotp$s (blue line) versus time for three opening holes in a 52\% volume fraction suspensions, shaken at $\Gamma=28$ and $f=45$Hz. The three curves have been time-shifted in order of increasing viscosity. Cleary, for increasing $\mu$ the (average) growth rate decreases.
}
\label{pic:NNopeningholes}
\end{figure}

In Fig.~\ref{pic:NNopeningholes} we show the typical time evolution of the hole diameter for a growing hole, here in a suspension consisting of the $\sigma=40$ $\mu$m particles and glycerol-water mixtures of three different viscosities. We observe that a lower viscosity causes holes to open faster. This appears to be comparable to the Newtonian liquids, where holes close faster for lower viscosity, but one needs to be careful in making this comparison: First of all, the viscosity of the suspending liquid is generally not comparable to the (non-constant) viscosity of the suspension, since there is a usually non-negligible or even dominant contribution from the particle phase. Secondly, we are now looking at the rate at which the hole grows \emph{against both} gravity and the suspension viscosity, whereas for the closing holes in a Newtonian liquid gravity was the driving force of the closure. This trend holds for all experiments we performed. Noteworthy is that for the higher suspending liquid viscosities and larger particles we typically observe growth of the hole until it develops a kink (where part of the system, including part of the wall, falls dry and remains separated by a steep slope (kink) from the rest of the system) whereas for small values of the suspending liquid viscosity and large particles we also observed holes that would go through many consecutive cycles of growth followed by a rapid collapse to an almost zero radius.

\begin{figure}[tbp]
\centering
\subfigure[]
{
\includegraphics[width=0.9\linewidth]{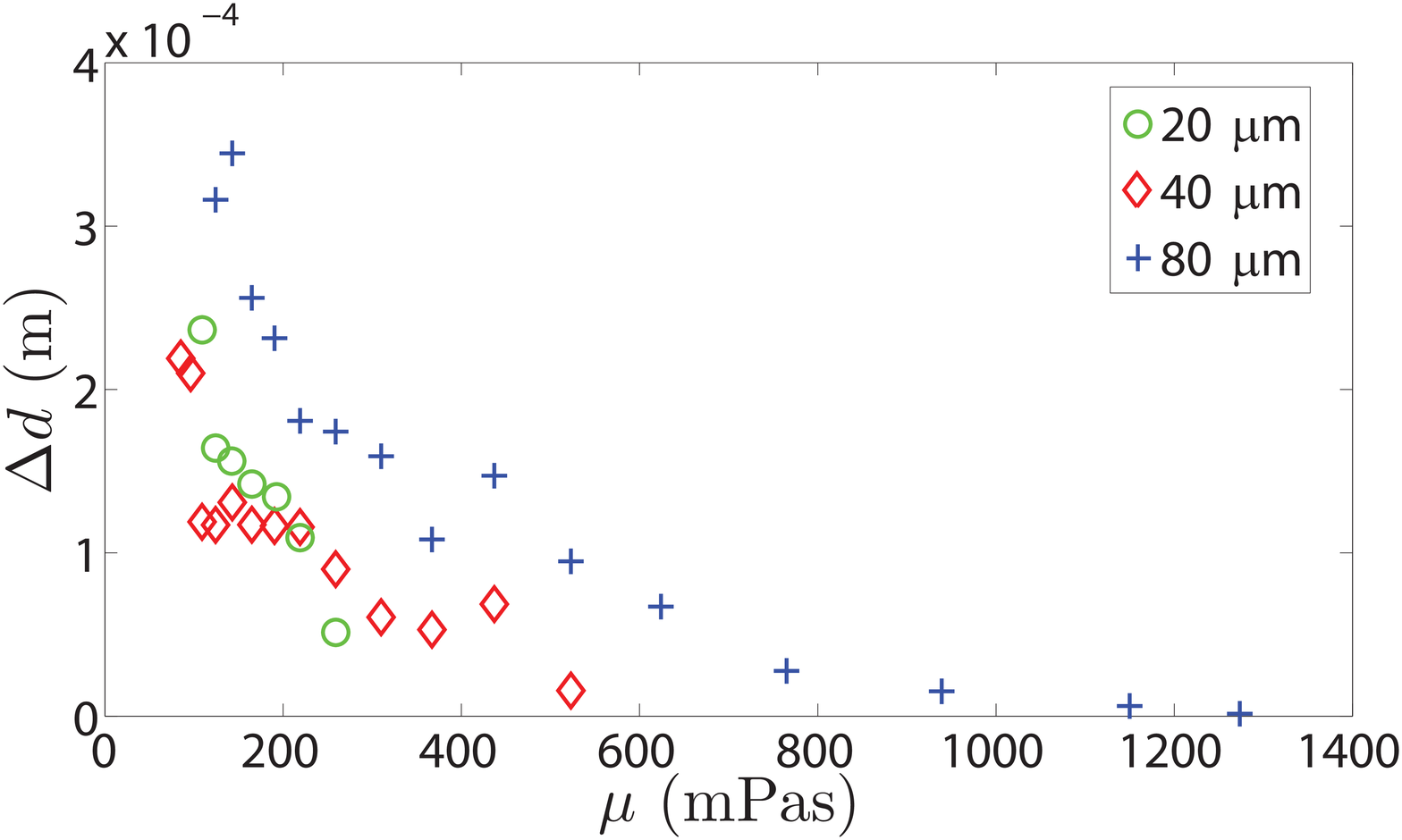}}
\subfigure[]
{
\includegraphics[width=0.9\linewidth]{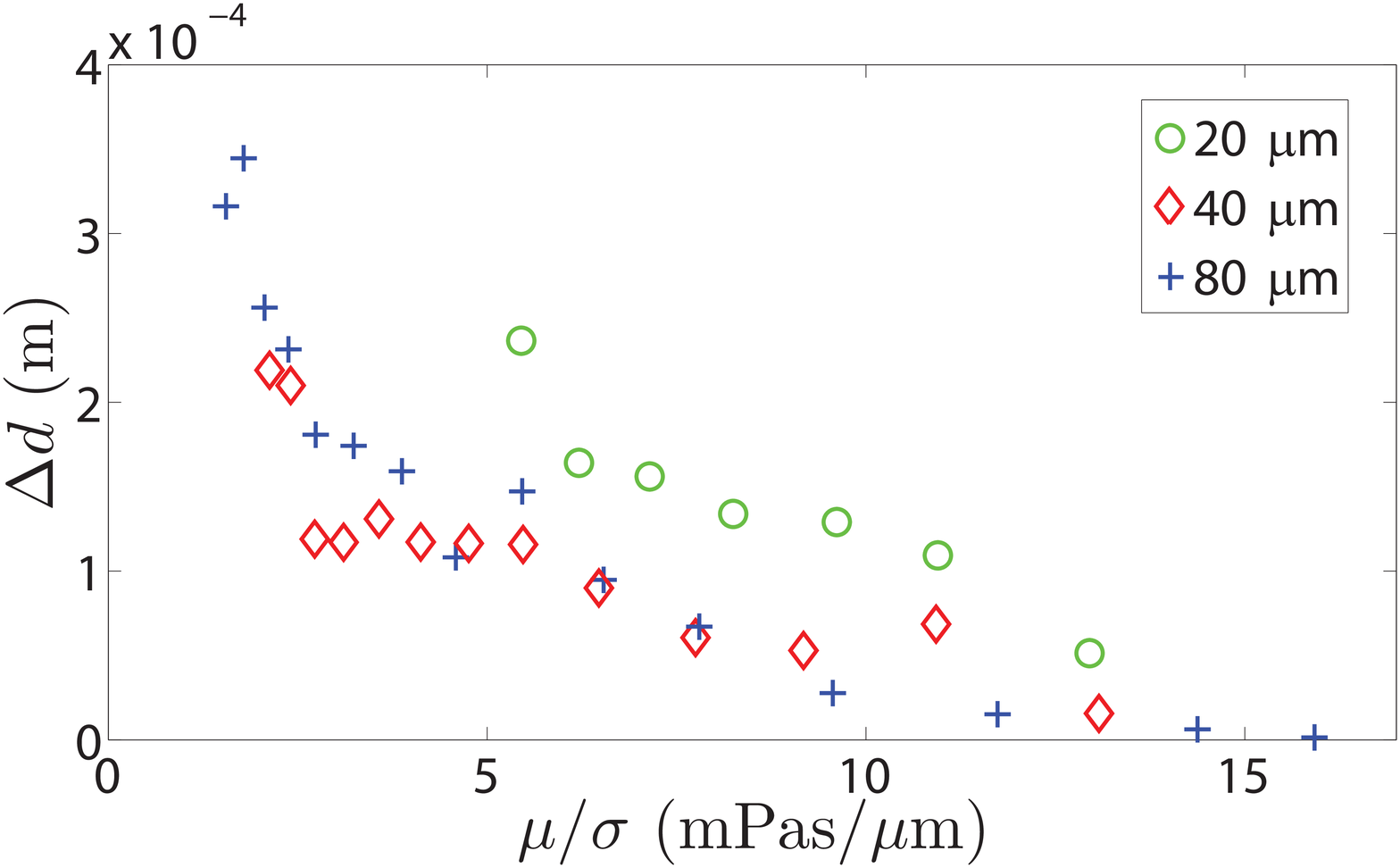}}
\caption{(a) The average growth $\Delta d$ per cycle of a growing hole as a function of the suspending liquid viscosity $\mu$. The experiments were done for suspensions of all three bead diameters, $\sigma = 20$ $\mu$m (green circles), $\sigma = 40$ $\mu$m (red diamonds), and $\sigma = 80$ $\mu$m (blue plusses) and a packing fraction of $\phi = 0.52$. The driving parameters are $f=45$ Hz and $\Gamma = 28$. (b) The same data as in (a) but now plotted as a function of $\mu/\sigma$, the suspending liquid viscosity over the particle diameter.}
\label{pic:NNopening}
\end{figure}

To further quantify the dependence of the growth rate on the suspending liquid viscosity, we determined the average growth $\Delta d$ of the hole diameter per cycle and plotted it against the suspending liquid viscosity $\mu$ in Fig.~\ref{pic:NNopening}(a) for all three bead sizes. We observe that all three data sets show a clear decrease of $\Delta d$ with increasing $\mu$, confirming our observation that the growth rate decreases with increasing suspending liquid viscosity. The data however does not collapse onto a single curve. Therefore, in Fig.~\ref{pic:NNopening}(b) we plot the same data as a function of $\mu / \sigma$ which leads to a reasonable collapse of the data for the two larger sizes, the significance of which will be discussed further down.

\begin{figure}[tbp]
\centering
\includegraphics[width=0.99\linewidth]{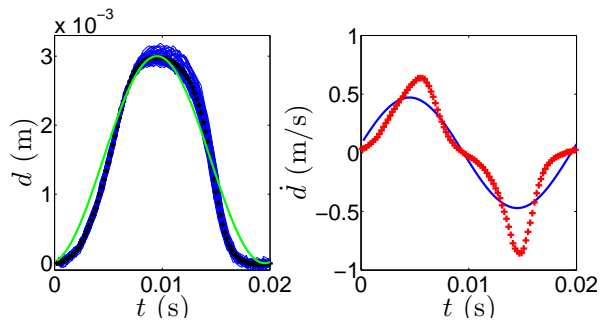}
\caption{(a) Overlay of many cycles of a growing hole experiment with $40$ $\mu$m beads suspended in a glycerine-water mixture of viscosity $\mu=0.52$ Pa$\cdotp$s, all shifted to start at $t=0$ and the initial diameter shifted to $d=0$ (blue curves). The black plus symbols indicate the cycle-averaged hole diameter. 
The yellow line is a sine fit through the average (thus neglecting the actual growth of the hole). (b) The instantaneous velocity of the growing hole averaged over all cycles (red plus symbols). The solid (blue) line is the derivative of the sine fit of (a). The experimental parameters are, $h_0=6$mm, $f=45$ Hz, $\Gamma = 28$, and $\phi = 0.52$.
}
\label{pic:NNsymmetry}
\end{figure}

Just like we have done for the Newtonian liquids (cf. Fig.~\ref{pic:Nsymmetry}), we can overlay many single cycles and compute the cycle-averaged diameter and velocity, the result of which is plotted in Fig.~\ref{pic:NNsymmetry}. This reveals several prominent features: The first is that --quite unlike for the closing holes in the Newtonian liquids-- the signal deviates significantly from a sinusoidal shape. This is especially clear when comparing the cycle-averaged velocity to the derivative of the sine fit [Fig.~\ref{pic:NNsymmetry}(b)]. In this plot we find a second remarkable feature: The magnitude of the most negative velocity ($\dot{d} \approx -0.90$ m/s) is larger than that of the most positive velocity ($\approx 0.65$ m/s), which is surprising since the hole on average is growing, i.e., for the time average we have $\langle\dot{d}\rangle>0$. When determining the duration of the opening and closing parts of the cycle, we find that they tend to lie very close to one another, implying that large closing velocities occur in a narrow time interval, whereas large opening velocities are found in a broader period of time. Indeed, the second, closing half of the cycle is sharply peaked, compared to a wider, somewhat closer to sinusoidal, shape during the opening half. This is clearly observed in Fig.~\ref{pic:NNsymmetry}(b).

\begin{figure}[tbp]
\centering
\subfigure[]
{
\includegraphics[width=0.9\linewidth]{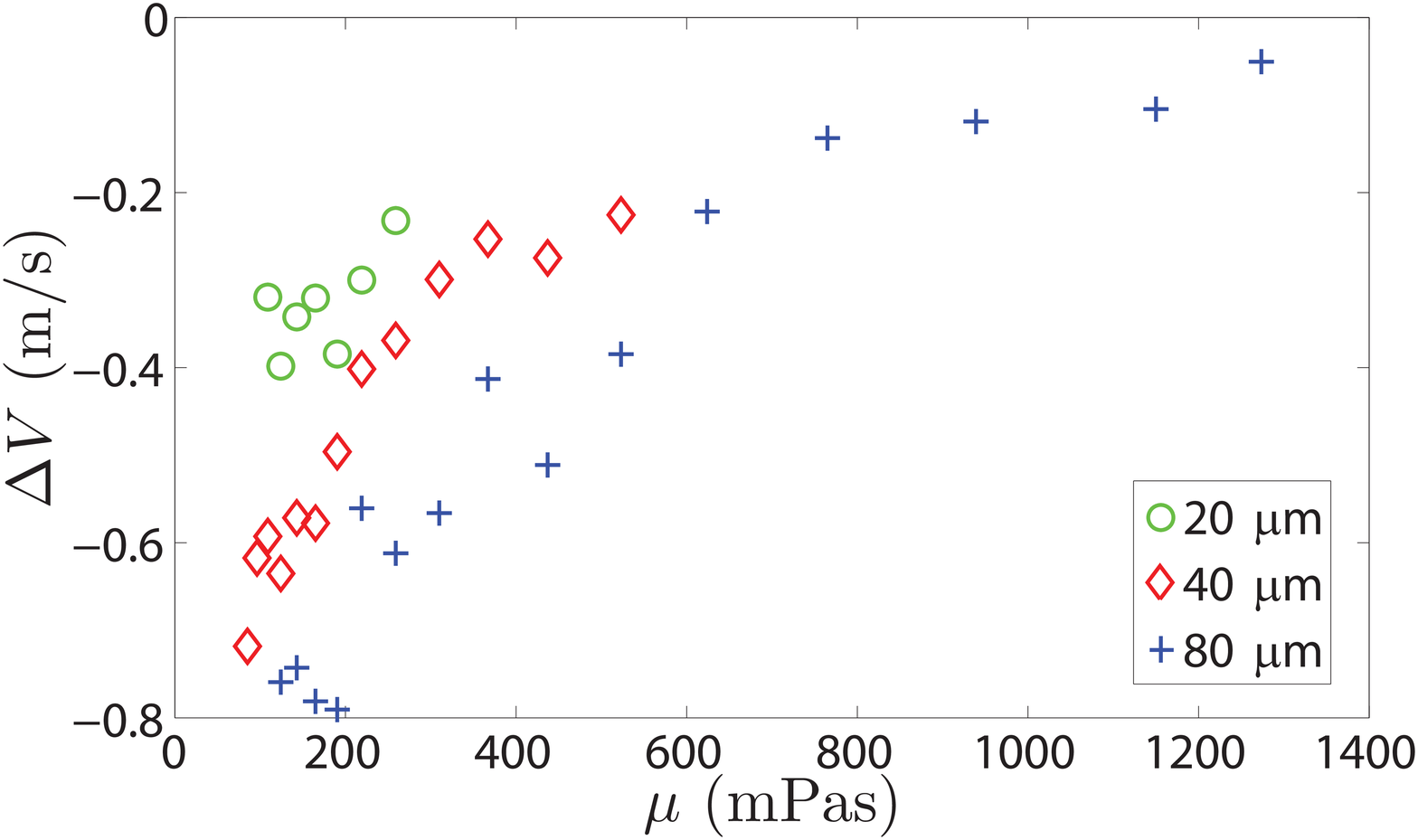}}
\subfigure[]
{
\includegraphics[width=0.9\linewidth]{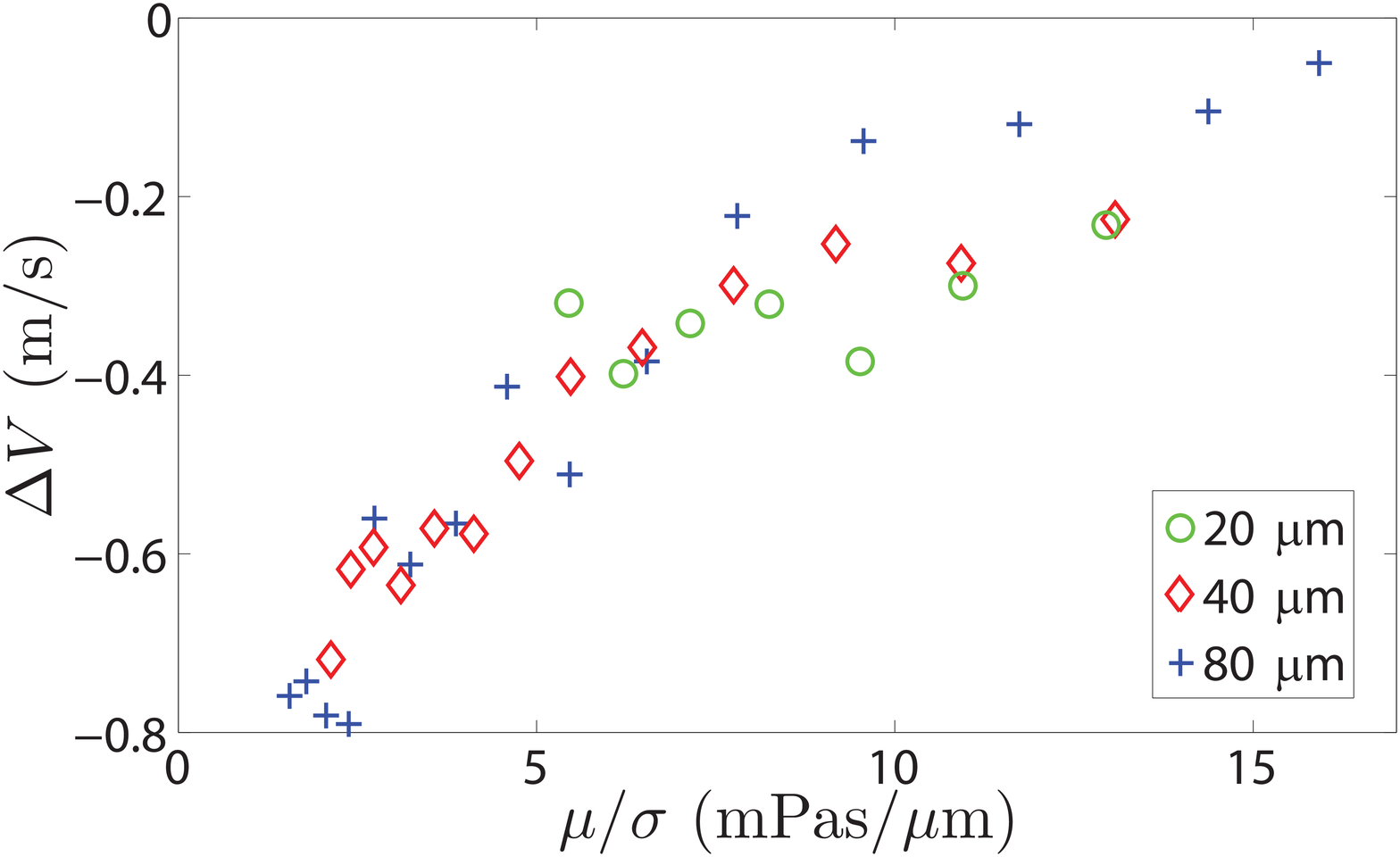}}
\caption{(a) The difference $\Delta V \equiv \max(\dot{d}) + \min(\dot{d})$ in the maximum opening and closing velocities in the growing hole state, averaged over all cycles as a function of the suspending liquid viscosity $\mu$, again for suspensions of all three bead diameters, $\sigma = 20$ $\mu$m (green circles), $\sigma = 40$ $\mu$m (red diamonds), and $\sigma = 80$ $\mu$m (blue plusses) and a packing fraction of $\phi = 0.52$. As before, the driving parameters are $f=45$ Hz and $\Gamma = 28$. (b) The same data as in (a) but now plotted as a function of $\mu/\sigma$.
}
\label{pic:NNdeltaV}
\end{figure}

The above asymmetry is visible in all of our experiments, as can be seen in Fig.~\ref{pic:NNdeltaV}, where we plot the difference between the magnitudes of the largest opening and closing velocities $\Delta V \equiv |\max(\dot{d})| - |\min(\dot{d})| = \max(\dot{d}) + \min(\dot{d})$. The fact that $\Delta V$ is always negative expresses that the magnitude of the most negative velocity is larger than that of the most positive. Just like the average growth $\Delta d$ per cycle decreased with increasing viscosity, so does the magnitude of the velocity difference $\Delta V$, which becomes less negative as $\mu$ becomes larger. In addition we find that the data for the different hole sizes are rather scattered in the $\Delta V$ versus $\mu$ plot, but appear to collapse when plotted against $\mu/\sigma$.

\begin{figure}[tbp]
\centering
\includegraphics[width=0.9\linewidth]{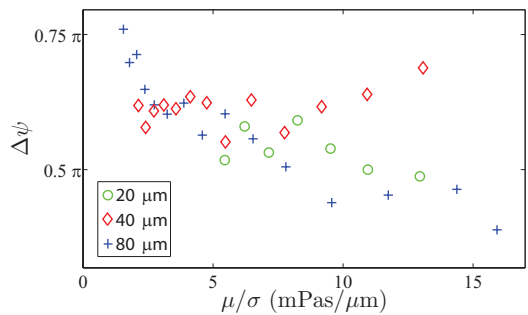}
\caption{The phase difference $\Delta\psi$ between the vertical position of the container (vibrated at $f=45$ Hz and $\Gamma = 28$) and the diameter of a growing hole, averaged over all cycles. This quantity is plotted as a function of the suspending liquid viscosity $\mu$, again for suspensions of all three bead diameters, $\sigma = 20$ $\mu$m (green circles), $\sigma = 40$ $\mu$m (red diamonds), and $\sigma = 80$ $\mu$m (blue plusses) and a packing fraction of $\phi = 0.52$.
}
\label{pic:NNphase}
\end{figure}

Finally, we can determine the phase shift $\Delta\psi$ between the driving and the hole although this is slightly more difficult than in the Newtonian liquid case (as well as suffering from some arbitrariness) because of the deviations from the sinusoidal shape. The results are plotted as a function of $\mu/\sigma$ in Fig.~\ref{pic:NNphase}: Again the horizontal oscillation of the hole lags behind the vertical container position but now by a phase shift that is slightly larger than $\pi/2$ and that increases somewhat when the viscosity of the suspending liquid becomes smaller or the particle size becomes larger. So again the velocity with which the hole oscillates is in antiphase with the container acceleration, and as a consequence the hole velocity is in phase with the acceleration of the suspension layer.

\subsection{Interpretation}\label{suspensionmod}

It is now time to make an inventory of what we believe happens when we create a hole-shaped disturbance in a liquid layer in  a container which is oscillated vertically:
\begin{itemize}
\item{For highly viscous fluids (even if not Newtonian), the velocity of the hole walls is in phase with the acceleration the liquid layer experiences.}
\item{A viscous Newtonian liquid follows the acceleration perfectly, i.e., for a sinusoidal acceleration also the velocity is sinusoidal. This stands to reason since for a viscous fluid forcing (acceleration) and the response of the liquid (the velocity profile in the layer) are proportional, with viscosity $\mu$ as the proportionality constant (Section~\ref{Newtonian}).}
\item{Consequently, if the liquid is non-Newtonian the proportionality factor itself depends on the forcing and therefore the response of the liquid to a sinusoidal acceleration is a deformed signal. However, if stress depends monotonously on strain rate (like, e.g., in a power-law fluid) the deformation will be symmetric, i.e., sinusoidal with a superposition of only odd higher harmonics.}
\item{For our vertically vibrated suspension layers we find a non-symmetric velocity cycle. The second, negative velocity part is strongly deformed, whereas the first, positive velocity half is closer to sinusoidal~[Fig.~\ref{pic:NNsymmetry}(b)]. It appears that during the second half of the cycle the suspension behaves strongly non-Newtonian\footnote{In fact, from the shape of the curve in Fig.~\ref{pic:NNsymmetry}(b) one can deduce that it behaves like a shear-thinning fluid.} whereas during the first half the response is closer to that of a Newtonian fluid.}
\end{itemize}

The behavior in this last point may be summarized by saying that the behavior of the liquid is highly hysteretic. This is in (qualitative) agreement with the phenomenological model proposed by Deegan~\cite{refDeegan10}, who argued that a hysteretic rheology would be necessary to explain the existence of stable or growing holes in a vertically vibrated liquid layer.

Now, let us speculate about what could cause the suspension to respond in this manner. In the second (i.e., closing) half of the driving the suspension layer experiences a downward acceleration and, consequently, the suspension layer will be pushed against the bottom of the container and when set in motion by the presence of the hole it will do so with the typical non-Newtonian (shear-thinning) behavior that characterizes suspensions. In the first (opening) half of the driving, inertia actually creates a low pressure between the layer and the container bottom. Now suppose that this pressure gradient would be able to displace the liquid slightly with respect to the particle phase such that a thin layer of liquid --with a thickness comparable to the particle diameter $\sigma$-- forms between the bottom and the suspension. Such a layer could act as a lubrication layer,i.e., during this first half period the layer would move on top of this layer and the entire velocity gradient would be in this thin layer of Newtonian liquid, i.e., it would be a shear band.

This in turn would explain why the suspension layer in the first half behaves closer to a Newtonian fluid, namely because this thin liquid layer is a Newtonian fluid. More specifically, if we balance the gravitational energy of the suspension layer and the dissipation in the lubrication layer we obtain
\begin{equation}
\rho_s g h \sim \mu \frac{\dot{d}}{\sigma} \quad\Rightarrow\quad \dot{d} \sim \rho g h \frac{\sigma}{\mu}\,,
\end{equation}
i.e., the velocity $\dot{d}$ in the second half of the driving would scale as $(\mu/\sigma)^{-1}$. This is consistent with the fact that many of the observables ($\Delta V$ and $\delta d$) that characterize the growth of the hole, show a better collapse when plotted against $\mu/\sigma$ rather than $\mu$ itself. Conversely, one could state that dependence on $\mu/\sigma$ indicates the existence of a shear layer of suspending liquid (with viscosity $\mu$) and thickness $\sim \sigma$.

Incidentally, the presence of such a thin shear layer may also account for the convection rolls that have been observed in the rim of these structures~\cite{refEbata09,reffalcon12}: In the second, closing half of the driving the suspension responds with a flow profile in the layer in with the largest velocity on top and zero velocity at the bottom. In the first, opening half the layer slides back as a whole, on top of the thin shear layer. Consequently, the displacement per cycle of a fluid element near the bottom is different from that near the top, giving rise to a convection roll.

\section{Conclusions}\label{conclusions}

We have comparatively studied the dynamics of holes in a vertically vibrated layer of viscous Newtonian liquids on the one hand and of dense particle suspensions on the other. We find that all the holes oscillate with a phase shift of $\Delta\psi = \pi/2$ with respect to the driving signal, such that the velocity of the hole is in phase with the vertical acceleration experienced by the fluid layer in the frame of reference of the container. In the Newtonian liquids we observe that holes always close, while in the suspensions holes may grow in time, depending on the driving parameters.

For the Newtonian liquids we find that the closing velocity is inversely proportional to the liquid viscosity, which is explained from a simple balance of gravitational and viscous forces. The presence of the driving is seen to hardly influence the closing: Independent of frequency and acceleration of the driving we find that the cycle-averaged closing rate of the holes is the same as for a closure that is driven by gravity only. We present a lubrication model for the closure of holes which is in good agreement with the experiments.

For the suspensions we focus on the growing holes regime and find that the growth rate of these growing holes is proportional to the ratio of the particle diameter and the suspending liquid viscosity. Comparing the growing holes to the closing ones in Newtonian liquids, we observe that in suspensions the response is highly non-linear. In addition, the symmetry of the oscillation is broken, with  larger inward velocities than outward ones, which is surprising since the hole is growing. The reason is that large outward velocities only occur in a small time interval, whereas the inward ones are spread over the entire half-period. We tentatively explain this asymmetry from the formation of a thin lubricating layer of suspending liquid between the suspension and the bottom in the first half-period in which the hole is opening.

The work is part of the research program of FOM, which is financially supported by NWO.

\appendix

\section{Modeling of hole closure in a viscous layer} \label{appendix}

A lubrication model of an axisymmetric viscous layer $h(r,t)$ starts with the axisymmetric Stokes' equation in the thin layer limit, with pressure given by the hydrostatic pressure in the layer $p=\rho g[h(r,t)-z]$. Neglecting gradients in the radial direction in comparison to those in the vertical direction we than integrate
\begin{equation}\label{eqA:velprofile}
	\mu\frac{\partial^2 u_r}{\partial z^2} = \frac{\partial p}{\partial r} \quad\Rightarrow\quad u_r = \frac{\rho g}{2\mu}\frac{\partial h}{\partial r} z(z-2h)\,,
\end{equation}
where we have used the no-slip boundary condition at the bottom ($u_r(0)=0$) and the free-slip condition at the free surface ($\partial u_r/\partial z (h) = 0$). Continuity, integrated over the layer height gives
\begin{equation}
	\frac{\partial h}{\partial t} = -\frac{1}{r}\frac{\partial}{\partial r} \left[ r\,\int_{0}^h u_r(z,t) dz\right]\,,
\end{equation}
which with Eq.~(\ref{eqA:velprofile}) immediately leads to Eq.~(\ref{eq:gravityonly})
\begin{equation}\label{eqA:gravityonly}
\frac{ \partial h}{ \partial t} = \frac{\rho g}{3 \mu r} \frac{\partial}{\partial r} \left[r h^3 \frac{ \partial h}{\partial r} \right]\,.
\end{equation}

If we look to compute the closure of a hole of initial diameter $d_0$ (at $t=0$ s) in an infinite layer of liquid of thickness $h_0$, we can find a similarity solution to Eq.~(\ref{eqA:gravityonly}). To find it, we first nondimensionalize $h$, $r$, and $t$ with the length and time scale in the problem, namely $h_0$ and $t_0 \equiv 3\mu/(\rho g h_0)$ respectively. If we now use a selfsimilar ansatz $\widetilde{h} = \widetilde{t}^\alpha H(\widetilde{r}/\widetilde{t}^\beta)$ in Eq.~(\ref{eqA:gravityonly}), we find a solution provided that $\alpha = 0$, $\beta=1/2$
\begin{equation}\label{eqA:solution}
h(r,t) = h_0\,\, H\!\!\left(\sqrt{\frac{3\mu r^2}{\rho g h_0^3(t_c - t)}}\right) \,\,,
\end{equation}
where $t_c$ is the time at which the hole closes and $H(\eta)$ is a solution of
\begin{eqnarray}\label{eqA:ODE}
\frac{d^2 H^4}{d\eta^2} \,\,+\,\, \frac{1}{\eta} \frac{dH^4}{d\eta} = -2\eta \frac{dH^4}{d\eta} \nonumber\\
H(\infty) = 1\,\,; \quad H(\eta_0)=0 \,\,.
\end{eqnarray}
The fact that $\eta_0$ needs to be a constant implies that the rim diameter $d(t)$ should scale as
\begin{equation}\label{eqA:rim}
d(t) = d_0 \sqrt{\frac{(t_c-t)}{t_c}}  = 2 \eta_0 \sqrt{\frac{\rho g h_0^3(t_c-t)}{3\mu}}\,\,.
\end{equation}
Note that the problem is not uniquely determined by providing $d_0$, and that in addition the closure time $t_c$ needs to be supplied to obtain a full solution to the problem. Then, $\eta_0$ can be determined as $\eta_0 = [3\mu d_0^2/(4\rho g h_0^3 t_c)]^{1/2}$ and Eq.~(\ref{eqA:ODE}) has a unique solution. Note, that in this case the initial profile is also fixed by the self-similar solution. Alternatively, one could therefore also start from the initial profile, match it to the solution of Eq.~(\ref{eqA:ODE}) for a certain $\eta_0$ which then fixes $t_c$. (This can be done provided that the initial profile is compatible with the equations.)
\bigskip

To obtain solutions of Eq.~(\ref{eqA:gravityonly}) that are more realistic given the experimental setup that we use we turn to numerical simulations. Here, we replace the actual boundary conditions at the side wall (zero radial velocity and no-slip) --which are impossible to incorporate into the lubrication model-- with the following integral statement of mass conservation in the system
\begin{equation}\label{eqA:masscons}
\int_{r=0}^{D/2} h(r,t) r dr = \text{constant}\,\,,
\end{equation}
where $D$ is the diameter of the container, or equivalently, taking the time derivative of Eq.~(\ref{eqA:masscons}) and using Eq.~(\ref{eqA:gravityonly})
\begin{equation}\label{eqA:massconsdiff}
\tfrac{1}{2} D \left[h^3\frac{\partial h}{\partial r}\right]_{r = D/2} \!\!\!\!\!\!= 0\quad\Rightarrow\quad \left.\frac{\partial h}{\partial r}\right|_ {r = D/2} \!\!\!\!\!\!= 0\,\,,
\end{equation}
where it was used that $h(D/2,t)>0$. Eq.~(\ref{eqA:gravityonly}) is of a type that is known as a non-linear diffusion equation, which is of a very stable type that renders them easy to solve numerically. The equations are therefore solved with a simple forward integration scheme which lead to results that compare well to the experiments (see Section~\ref{model}).
\bigskip

Actually it is conceptually straightforward to incorporate the driving into the equations, as the only thing one needs to do is to substitute the gravitational acceleration $g$ with $g + a(t)$ where $a(t)$ is the instantaneous acceleration of the container, $a(t) = a \omega^2 \sin \omega t = \Gamma g \sin \omega t$ (with $\omega =  2\pi f$). This however has enormous implications for the numerical solvability of the equations, since for $\Gamma > 1$ there exists a time interval in each cycle for which $g + a(t) < 0$. In this interval Eq.~(\ref{eqA:gravityonly}) becomes a non-linear diffusion equation with a negative diffusion coefficient, which is terribly unstable and consequently extremely difficult to solve numerically. For the current problem there exists a workaround however, for which we need some additional understanding of the equations first.
\bigskip

To this end let us first examine a modified Eq.~(\ref{eqA:gravityonly}) without gravity
 \begin{equation}\label{eqA:shakingonly}
\frac{ \partial h}{ \partial t} =  \Gamma \sin(\omega t) \,\frac{\rho g}{3 \mu r} \frac{\partial}{\partial r} \left[r h^3 \frac{ \partial h}{\partial r} \right]\,.
\end{equation}
Clearly, a solution to Eq.~(\ref{eqA:shakingonly}) must have the same periodicity as the driving, i.e., $h(r,t+T)=h(r,t)$. Now, the simplest form that such a solution could have is $h(r,t) = h_s(r) + A(r)\exp[i\omega t + \varphi(r)]$, which corresponds to neglecting non-linear effects in Eq.~(\ref{eqA:shakingonly}). Now $h_s(r)$ can be any profile that satisfies the non-driven Eq.~(\ref{eqA:shakingonly}), i.e., $\partial h_s/\partial t = 0$, which can be any well-behaved function of $r$. Inserting this form into Eq.~(\ref{eqA:shakingonly}) and linearizing leads to
\begin{equation}
\omega A(r) e^{i(\omega t + \varphi(r) + \pi/2)} =  \Gamma \,\frac{\rho g}{3 \mu r} \frac{\partial}{\partial r} \left[r h_s^3 \frac{ \partial h_s}{\partial r} \right]\,e^{i\omega t}\nonumber\,\,,
\end{equation}
which needs to hold for any $t$, leading to
\begin{eqnarray}
A(r)  &=&  \frac{\Gamma}{\omega} \,\frac{\rho g}{3 \mu r} \frac{\partial}{\partial r} \left[r h_s^3 \frac{ \partial h_s}{\partial r} \right]\,\,,\nonumber\\
\varphi(r) &=& - \frac{\pi}{2}\,\,.
\end{eqnarray}
with which
\begin{equation}\label{eqA:solshakingonly}
h(r,t) = h_s(r) \,\,+\,\, \frac{\Gamma}{\omega} \,\frac{\rho g}{3 \mu r} \frac{\partial}{\partial r} \left[r h_s^3 \frac{ \partial h_s}{\partial r} \right]\,\exp[i(\omega t - \pi/2)]\,,
\end{equation}

\begin{figure}[tbp]
\centering
\includegraphics[width=0.99\linewidth]{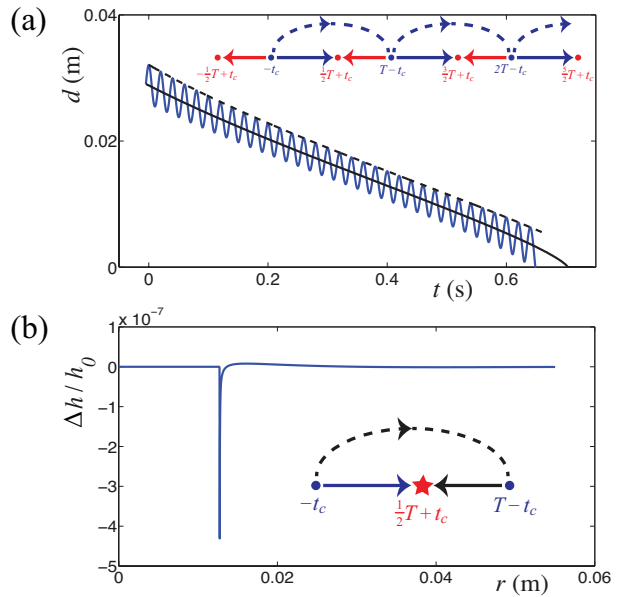}
\caption{(a) Time evolution of the hole diameter in a layer of a liquid of viscosity $\mu=1.23$ Pa$\cdotp$s, density $\rho = 1.26\cdot10^3$ kg/m$^3$ and thickness $h_0=6$ mm, driven at $f=50$ Hz and $\Gamma = 20$, obtained by numerically integrating Eq.~(\ref{eqA:full}) using the procedure described in the text (blue solid line). The black lines are solutions of the corresponding gravitational closure problem Eq.~(\ref{eqA:gravityonly}), one starting from the same initial condition as the blue line (solid) and the other is the one that is used in the integration procedure of the blue curve. The inset illustrates the integration procedure. (b) Difference between the profile at $t=T/2 + t_c$ obtained starting from $t=-t_c$ by direct integration of Eq.~(\ref{eqA:full}) and by integrating Eq.~(\ref{eqA:gravityonly}) to $t=T-t_c$ and integrating Eq.~(\ref{eqA:full}) backwards in time to $t=T/2 + t_c$. The inset illustrates this procedure.
}
\label{pic:appendix}
\end{figure}

The full equation, including both the driving and gravity, is equal to
 \begin{equation}\label{eqA:full}
\frac{ \partial h}{ \partial t} =  \left(\Gamma \sin(\omega t) + 1\right)\,\frac{\rho g}{3 \mu r} \frac{\partial}{\partial r} \left[r h^3 \frac{ \partial h}{\partial r} \right]\,.
\end{equation}
Note that, since we are dealing with $\Gamma \gg 1$, gravity is a small perturbation to Eq.~(\ref{eqA:shakingonly}). This implies that the gravitational timescale at which the profile decays ($t_g \sim 3\mu/(\rho g h_0)$) is typically much larger than that of the driving. I.e., if $h_g(r,t)$ is a solution to Eq.~(\ref{eqA:gravityonly}), on the timescale of a single period it does not significantly change and may hardly interfere with the oscillation. Using Eq.~(\ref{eqA:solshakingonly}), this suggests a solution to the full problem of the form
\begin{equation}\label{eqA:solfull}
h(r,t) = h_g(r,t) \,\,+\,\, \frac{\Gamma}{\omega} \frac{\partial h_g}{\partial t} \,\exp[i(\omega t - \pi/2)]\,,
\end{equation}
where we have used that $h_g(r,t)$ is a solution to Eq.~(\ref{eqA:gravityonly}) to simplify the expression for the amplitude of the oscillation.
\bigskip

This line of reasoning comes to the rescue when numerically solving Eqs.~(\ref{eqA:shakingonly}) and~(\ref{eqA:full}). A first useful trick is to realize that we can numerically integrate Eq.~(\ref{eqA:shakingonly}) from $t=0$ to $T/2$ since in this interval the coefficient of the right hand side is always positive. By going to a new time variable $\tau = -t$ we obtain a minus sign on the left hand side which exactly compensates for the minus sign of the coefficient in the interval $[-T/2,0]$. Consequently we can integrate Eq.~(\ref{eqA:shakingonly}) backwards in time from $t=0$ to $-T/2$, such that we obtain the solution on $[-T/2,T/2]$, i.e., a full cycle. Since the sought-for solution is periodic in time, this concludes our calculation.

For Eq.~(\ref{eqA:full}) we can proceed in a similar way and, by integrating both backwards and forwards from $t=-(T/2\pi)\,\arcsin(1/\Gamma) \equiv - t_c$ (where the coefficient changes sign), obtain a numerical solution on $[-T/2+t_c,T/2 + t_c]$, i.e., also on a full period of the driving. And, of course, it is impossible to extend this interval because it is bounded by an interval where the coefficient is positive on the negative side --such that backwards integration is not possible-- and similarly by an interval where the coefficient is negative on the positive side. We can however take the solution (cq. initial condition) $h(r,-t_c)$ and integrate it using Eq.~(\ref{eqA:gravityonly}), i.e., the equation that only contains gravity, from $-t_c$ to $T-t_c$. The solution $h_g(r,T-t_c)$ is now subsequently used as an initial condition for the full problem Eq.~(\ref{eqA:full}) which we then integrate on the interval $[T/2+t_c,3T/2 + t_c]$. If there is any truth in the analytical approximation Eq.~(\ref{eqA:solfull}) the two solutions should (approximately) match at the point where the two intervals meet, i.e., in $t=T/2+t_c$. This procedure can be iterated until the solution is obtained on the full time interval [see the inset of Fig.~\ref{pic:appendix}(a)].

In Fig.~\ref{pic:appendix}(a) we plot the result of this procedure for a layer of liquid with viscosity $\mu=1.23$ Pa$\cdotp$s and a thickness of $h_0=6$ mm, which is vibrated at a frequency of $50$ Hz and a dimensionless acceleration $\Gamma = 20$. The solid line is the solution of Eq.~(\ref{eqA:gravityonly}) starting from the same initial solution and the dotted line is the solution of that same equation that is used in the integration procedure. Clearly the integration procedure appears to work well. To quantify how good it actually works, in Fig.~\ref{pic:appendix}(b) we plot the difference between directly integrating from $t=-t_c$ to $t=T/2 + t_c$ and indirectly by using the gravitational solution to reach $t=T-t_c$ and integrating Eq.~(\ref{eqA:full}) backwards in time to $t=T/2 + t_c$. The difference, normalized by the initial layer thickness is never larger than $10^{-6}$, illustrating the accuracy of the procedure.

\end{document}